\documentclass[fleqn,usenatbib]{mnras}

\usepackage{newtxtext,newtxmath}

\usepackage[T1]{fontenc}
\usepackage{ae,aecompl}

\usepackage{graphicx}	
\usepackage{amsmath}	
\usepackage{amssymb}	

\newcommand{\OIII}{[O{\sc iii}]5007~\AA}

\newcommand \vhel{\ifmmode{~V_{{\rm HEL}}}\else{~$V_{{\rm HEL}}$}\fi}
\newcommand \kms{km s$^{-1}$}

\usepackage{color}
\usepackage[dvipsnames]{xcolor}
\definecolor{teal}{rgb}{0.0, 0.5, 0.5}

\title[The post-CE central star of PN NGC~2346]{On the post-common-envelope central star of the planetary nebula NGC~2346}

\author[A.~J. Brown et. al]{Alex J. Brown,$^{1,2}$\thanks{E-mail:
ajbrown2@sheffield.ac.uk} David Jones,$^{3,4}$\thanks{E-mail:
djones@iac.es} Henri~M.~J. Boffin$^5$,
and Hans Van Winckel$^6$
\\
$^{1}$Department of Physics and Astronomy, University of Sheffield, Sheffield, S3 7RH, UK\\
$^{2}$Isaac Newton Group of Telescopes, Apartado de Correos 368, E-38700 Santa Cruz de La Palma, Spain\\
$^{3}$Instituto de Astrof\'isica de Canarias, E-38205 La Laguna, Tenerife, Spain\\
$^{4}$Departamento de Astrof\'isica, Universidad de La Laguna, E-38206 La Laguna, Tenerife, Spain\\
$^{5}$European Southern Observatory, Karl-Schwarzschild-Str 2, 85748 Garching, Germany\\
$^{6}$Instituut voor Sterrenkunde, KU Leuven, Celestijnenlaan 200D bus 2401, 3001 Leuven, Belgium\\
}

\date{Accepted XXX. Received YYY; in original form ZZZ}

\pubyear{2018}

\begin{document}
\label{firstpage}
\pagerange{\pageref{firstpage}--\pageref{lastpage}}
\maketitle

\begin{abstract}
The common-envelope phase is one of the most poorly understood phases of (binary) stellar evolution, in spite of its importance in the formation of a wide-range of astrophysical phenomena ranging from cataclysmic variables to cosmologically important supernova type \textsc{i}a, and even recently discovered gravitational wave producing black hole mergers.  The central star of the planetary nebula NGC~2346 has long been held as one of the longest period post-common-envelope systems known with a published period of approximately 16 days, however the data presented were also consistent with much shorter periods of around 1 day (a more typical period among the known sample of post-common-envelope binary central stars).  Here, using the modern high-stability, high-resolution spectrograph HERMES, we conclusively show the period to, indeed, be 16 days while also revising the surface gravity to a value typical of a sub-giant (rather than main-sequence) resulting in an intrinsic luminosity consistent with the recently published GAIA parallax distance.  Intriguingly, the implied mass for the secondary ($\gtrsim$3.5 M$_\odot$) makes it, to our knowledge, the most massive post-common-envelope secondary known, whilst also indicating that the primary may be a post-RGB star.
\end{abstract}




\section{Introduction}

Central star binarity is now thought to be a key ingredient in understanding the formation and evolution of a large fraction of planetary nebulae \cite[PNe;][]{jones17c} - playing an important role in the observed morphologies \citep{hillwig16}, chemistry \citep{wesson18}, and perhaps even in the planetary nebula luminosity function \citep[PNLF; ][]{ciardullo05,davis18}.  However, very little is known about the processes by which binary stars can produce a PN - particularly the common envelope (CE) phase \citep[see e.g. the review of][]{ivanova13}.  One particularly interesting puzzle is the observed period distribution of post-CE central stars which shows a strong propensity of periods of a few days or less \citep{jones17c}, while models of the CE phase generally predict many more systems at longer periods \citep[see section 4 of][]{demarco08}.  As such, the properties of the few longer-period systems known, including recent discoveries by \citet{manick15}, \citet{sowicka17b} and \citet{miszalski18}, are of particular interest - likely holding the key to resolving this disagreement and, perhaps, even to understanding the CE phase itself.

The binary central star of NGC~2346 was one of the first to be discovered \citep{mendez78}, with a subsequent radial velocity study by \citet{mendez81} deriving an orbital period of roughly 16 days making it one of the longest period post-CE binaries known to-date.  However, there was considerable confusion over the true orbital period, with periods of around 1 day also presenting a reasonable fit to the data.  Later photometric studies also found a dominant 16 day period but in this case not directly attributable to the orbital motion of the binary but rather due to (variable) occultations of the binary by a dust cloud \citep{mendez82, acker85}.  Further support for such a long period comes from the observed nebular chemistry with \citet{wesson18} finding that PNe with shorter period binary central stars typically show extreme abundance discrepancies, while they place an upper limit on the abundance discrepancy factor of NGC~2346 of less than 10 (more consistent with a longer period central star).

As well as constraining the observed radial velocity variability, \citet{mendez81} also derived the spectral type of the secondary star in the system (the hot primary is not visible in their optical spectra) concluding it to be an A-type star of mass M=1.8 M$_\odot$, temperature T$_\mathrm{eff}$=8000 K and surface gravity log g=4.00.  These values imply a distance of 0.7$\pm$0.1 kpc to the system, however the parallax as measured by GAIA result in a much larger distance of 1.45$^{+0.09}_{-0.08}$ kpc \citep{gaiadr2}.  This larger distance is also consistent with the distance to the nebula (D=1.57$\pm$0.54 kpc) as derived using the H$\alpha$ surface brightness - radius relation of \citet{frew16}.

In this paper, we present a study, based on VLT-FORS2 and Mercator-HERMES spectroscopy, of the central star of NGC~2346 to revisit the orbital period and stellar parameters in order to critically re-evaluate its status as one of the longest period post-CE binaries known as well as try to reconcile the apparent discrepancy between parallax distance and the distance implied by previous modelling attempts.

\section{Radial velocity monitoring}
\label{sec:hermes}

The central star of NGC~2346 was observed 33 times (1800-s exposure time), between 2016 November 27 and 2018 April 18, using the Hermes Spectrograph mounted on the 1.2-m Mercator Telescope at the Observatorio del Roque de Los Muchachos on the Spanish island of La Palma \citep{raskin11}.  The pipeline reduced data were then continuum subtracted using the \textsc{iSpec} software \cite{blanco-cuaresma14} and cross-correlated against an A-type spectral template produced using \textsc{spectrum} \citep{spectrum}.  The resulting heliocentric radial velocity measurements are shown in table \ref{tab:rvs}.

\begin{figure}
\centering
\includegraphics[width=\columnwidth]{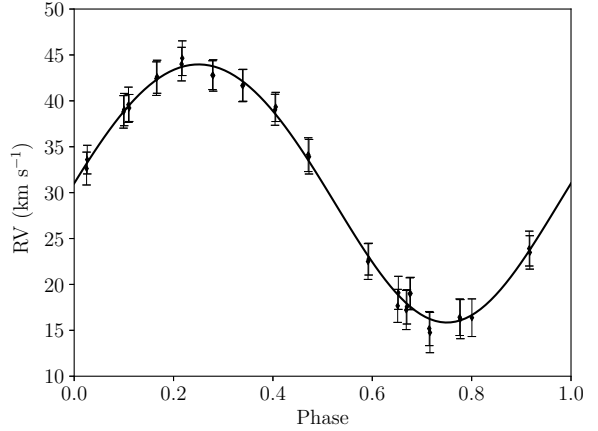}
\caption{Radial velocity curve of NGC~2346}
\label{fig:rvcurve}
\end{figure}

\begin{figure}
\centering
\includegraphics[width=\columnwidth]{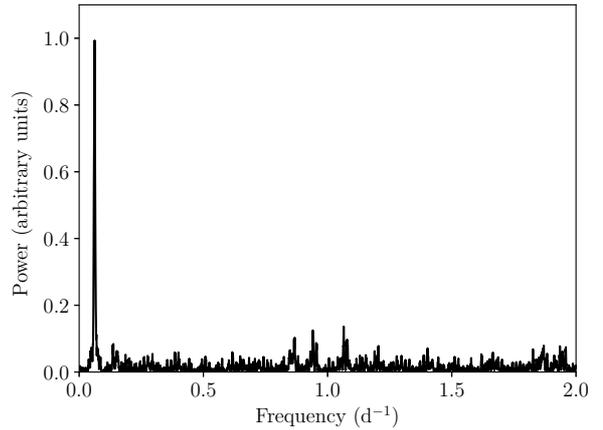}
\caption{Power spectrum of the radial velocity observations of NGC~2346 showing the clear peak at a frequency of 0.0625 d$^{-1}$ (P=16d).}
\label{fig:powerspec}
\end{figure}

The radial velocities were then fit using the \textsc{radvel} package \citep{radvel}, sampling the posterior probability densities of the orbital parameters (period, semi-amplitude, eccentricity, argument of periastron) via Markov Chain Monte Carlo.  The data are shown folded on the resulting best fit solution in figure \ref{fig:rvcurve} while the parameters of the fit and their uncertainties are listed in table \ref{tab:rvparams}.  Fits were attempted forcing shorter orbital periods \citep[with P$\sim$1d consistent with the possible periodicities identified by][]{mendez81}, but in all cases the quality of the fit was significantly poorer than for a period of 16 days. In figure \ref{fig:powerspec}, we present a \textsc{clean}ed power spectrum, computed via ten iterations with a loop gain of 0.1 \citep{roberts87}, of the radial velocity observations clearly showing that the 16d period derived by the fit is the only convincing peak.
As such, the new Mercator-HERMES data confirm the period favoured by \citet{mendez81}, fully ruling out the shorter period aliases which contaminated their data. We have also verified that adding the data from \citet{mendez81} did not improve the orbit, given the very high quality data from Hermes. 

\begin{figure}
\centering
\includegraphics[width=\columnwidth]{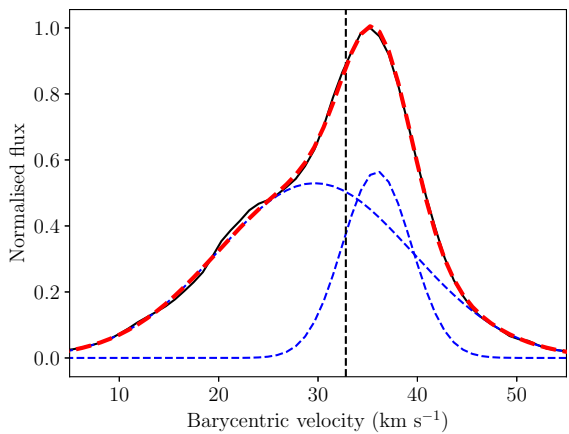}
\caption{Gaussian fits of the \OIII{} emission from the PN NGC~2346.}
\label{fig:oiii}
\end{figure}

The systemic velocity as measured by the fit ($\gamma$=30.5$\pm$0.8 \kms{}) is appreciably different from the values quoted in the literature for the nebula \citep[with most lying in the range 20--25 \kms{};][]{durand98,mendez81}.  However, the systemic nebula velocity, as measured by fitting two gaussians (consistent with the two features arising from the front and back ``walls'' of the bipolar structure) to the \OIII{} emission line profile and taking the centroid (see figure \ref{fig:oiii}), of our observations is found to be marginally consistent with the stellar systemic velocity ($v_{neb}$=32.7$\pm$1.6 \kms{}).  \citet{arias01} provide support for a similar nebular velocity, as they find that the maximum intensity of their Fabry-Perot scans lies on the channel map at 27 \kms{}, while of the channel maps either side of this the redder map (at 37 \kms{}) is more symmetrical than its blue counterpart (at 17\kms{}) - indicative that the true systemic velocity lies between 27 and 37 \kms{} just as our measured values for both nebula and binary do.

\begin{table}
\caption{Parameters of the central star of NGC~2346 derived from the Mercator radial velocities}
\label{tab:rvparams}
\begin{tabular}{rl}
\hline
Orbital period, P & 16.00$\pm$0.03 d\\
Systemic velocity, $\gamma$ & 30.5$\pm$0.8 \kms{}\\
Semi-amplitude, K & 14.1$\pm$0.6 \kms{}\\
Eccentricity, e & 0.04$\pm$0.03\\
Argument of periastron, $\omega$ & 180\degr{}$\pm$160\degr{}\\ 
Binary mass function, $f$ & 0.00464$\pm$0.00062 M$_\odot$\\
\hline
Nebular systemic velocity & 32.7$\pm$0.6 \kms{}\\
\hline
\end{tabular}
\end{table}

\section{Stellar parameters and abundances}

The central star of NGC~2346 was observed using the FORS2 instrument of the ESO VLT's UT1 Antu telescope \citep{FORSshort} with single spectra taken back-to-back using the 1200B (600-s exposure time) and 1200R (along with the GG435 order blocking filter, 60-s exposure time) grisms.  A 0.7\arcsec{} slit was employed with the MIT/LL CCD mosaic binned 2$\times$2 ($\equiv$0.25\arcsec{} per binned pixel) to provide a spectral resolution of 1--2\AA{} across the observed wavelength range $\lambda$$\sim3600-5000$\AA{} (1200B grism) and $\lambda$$\sim5800-7200$\AA{} (1200R grism).  The spectra were bias-subtracted, wavelength-calibrated and flux-calibrated using bias, arc lamp and standard star observations acquired as part of ESO's standard calibration plan.  The spectra were then sky-subtracted (using only regions dominated by nebular emission, in order to subtract both sky and nebular contributions) before optimal extraction.

\begin{figure}
\centering
\includegraphics[width=\columnwidth]{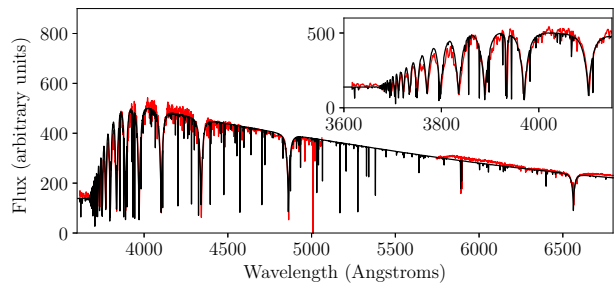}
\caption{FORS2 spectrum of the central star of NGC~2346 (red) along with the best-fitting synthetic spectrum (black) implying an effective temperature, T$_\mathrm{eff}$=7750 K, and a surface gravity, log g=3.0.}
\label{fig:fors2spec}
\end{figure}

\begin{figure}
\centering
\includegraphics[width=\columnwidth]{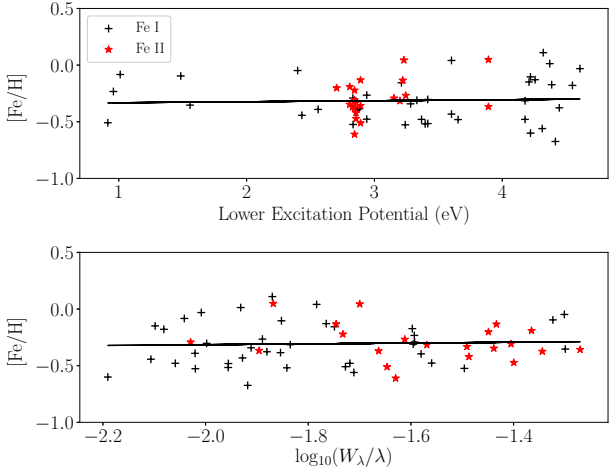}
\caption{Iron abundances in the secondary component of the central star of NGC~2346 as a function of excitation potential (upper panel) and reduced equivalent width (lower panel).}
\label{fig:equivwidths}
\end{figure}

\begin{figure}
\centering
\includegraphics[width=\columnwidth]{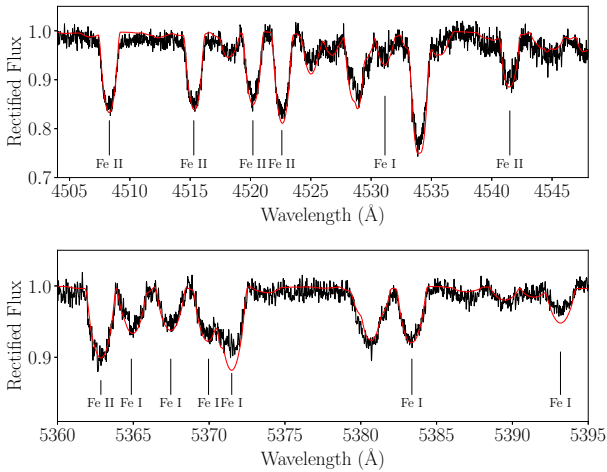}
\caption{Observed (black) and synthetic (red) spectrum shown around two regions containing multiples lines of singly and doubly ionized iron, highlighting the quality of the fit. The synthetic spectrum was produced using parameters T$_\mathrm{eff}$=7750 K, log g=3.0 and [Fe/H]=-0.35}
\label{fig:ironlines}
\end{figure}

The temperature and surface gravity of the star were probed by comparing the resulting extracted spectra to synthetic spectra produced by \textsc{spectrum} using model atmospheres from \citet{ck2004} which had been reddened assuming E(B-V)=0.25 and R$_v$=3.1 \citep{frew16}.  The best-fitting model (shown overlaid on the observed spectrum in figure \ref{fig:fors2spec}) was found to have an effective temperature T$_\mathrm{eff}$=7750$\pm$200 K and surface gravity log g=3.0$\pm$0.25. Consistent values were derived using the equivalent widths of iron lines present in the Mercator-Hermes spectrum presented in section \ref{sec:hermes}.  Furthermore, this analysis allowed the derivation of the metallicity of the companion resulting in a best-fitting value for [Fe/H]$\sim-0.35$  (see figure \ref{fig:equivwidths}).  The quality of the fit is demonstrated for a selection of singly and doubly ionized Iron lines in figures \ref{fig:ironlines}.  Lines of many other elements were present in the spectra including the s-process elements barium and strontium, and several elements from groups 2 (calcium, magnesium), 3 (scandium, yttrium) and 4 (titanium, zirconium).  These were all probed for signs of inconsistency with respect to the derived metallicity \citep[perhaps as a result of chemical contamination from the primary around the time of the CE phase, e.g. ;][]{miszalski13b}, however no evidence for deviation in their abundances from those expected for the measured metallicity were found.

Collectively, the derived stellar parameters (listed in table Table \ref{tab:modpars}) are consistent with evolutionary tracks of a sub-giant star of mass $\sim$3.5 M$_\odot$ \citep{bertelli09}. These tracks are extremely dependent on a multitude of factors, however a further sanity check of this mass is provided by the GAIA parallax distance (D=1.45 kpc).  The luminosity of such a star (log(L/L)$_\odot\sim$2.4) implies an un-extincted apparent magnitude of approximately 10.8 at a distance of 1.45 kpc \citep[assuming a typical bolometric correction of -0.2;][]{pickles98} - which is roughly consistent with the measured extinction corrected magnitude of \citet{kohoutek95} at 10.9 mag, particularly when allowing for the effects of the observed large amplitude variability \citep{kohoutek83} and accounting for the uncertainties on the measured parameters (gravity, effective temperature, metallicity and extinction) and evolutionary tracks.

\begin{table}
\caption{Atmospheric parameters for the \textsc{spectrum} model}
\label{tab:modpars}
\begin{tabular}{rll}

\hline
Effective temperture, T$_\mathrm{eff}$ & 7750 & $\pm$ 200 K \\
Surface gravity, log(g) & 3.0 & $\pm$ 0.25 dex\\
Metallicity, $[$Fe/H$]$ & -0.35 & $\pm$ 0.2 dex\\
Microturbulence, $\xi$ & 4.0 & $\pm$ 0.1 \kms{} \\
Rotational velocity, vsin$i$ & 52 & $\pm$ 5 \kms{} \\
\hline

\end{tabular}
\end{table}

\section{Discussion}

We have conclusively shown that the orbital period of the binary central star of NGC~2346 is, as previously derived by \citet{mendez81}, approximately 16 days.  Furthermore, detailed study of the stellar parameters via spectral synthesis revises the secondary in the system from a main-sequence star to a sub-giant, the absolute magnitude of which is consistent with the recently derived GAIA parallax distance.

At an inclination of 65$^\circ$ \citep[consistent with the binary plane being aligned with the waist of the bipolar nebula;][]{arias01}, and assuming a canonical mass of 0.6 M$_\odot$ for the primary, the binary mass function (derived from the Hermes radial velocity curve) implies a secondary mass of $\sim$5.3 M$_\odot$.  For the derived surface gravity and temperature, such a massive companion would present with a luminosity inconsistent with the distance derived by GAIA.  As such, either the binary plane is not aligned with the nebular waist (a lower inclination would imply a lower secondary mass), or the primary has a lower mass.  The former option seems unlikely given that, in all cases where both are known, the binary orbital inclination is found to be coincident with the nebular waist - with a probability of chance a alignment being less than one in one million \citep{hillwig16}.  The orbital period of NGC~2346 is significantly longer than the other systems, however all are post-CE meaning that one would not expect a misalignment\footnote{Some evidence for misalignment between binary and nebula has been found in LoTr~5 \citep{jones17b,aller18}.  However, the orbital orbital period of the central star of LoTr~5 is much longer ($\sim$2700 d), clearly differentiating its evolution from the others considered in that it has not experienced a CE phase.}.  It is, perhaps possible that the quoted inclination of the nebula is not accurate, with a value of $\sim$45$^\circ$ bringing the secondary mass down to 3.5 M$_\odot$ (consistent with the derived atmospheric parameters and evolutionary tracks).

The second possibility, that the nebular progenitor has a lower mass, is perhaps equally unlikely.  In order to be the more evolved component of the binary, the initial mass of the primary must have been greater than that of the secondary which, given that we do not see any evidence for chemical contamination in its atmosphere, is likely to be very similar to its current mass\footnote{It is possible that the mass transfer could have occurred before the primary could become significantly chemically enriched (i.e. while it was still on the RGB), which would not result in chemical enrichment of the secondary.  However, other post-CE systems, where significant accretion from primary to secondary has occurred, indicate that the majority of the mass transfer occurs very shortly before entering the CE phase \citep{miszalski13b,jones15} - meaning that if there was significant transfer of non-chemically enriched material while the primary was on the RGB it is also likely that the system experienced the CE while the primary was still on the RGB.}.  The white dwarf initial-final mass relation would imply that the primary's mass should be $\sim$0.8 M$_\odot$ \citep[e.g.;][]{elbadry18}, unless its evolution was cut at a particularly early stage by the CE.  For example, assuming a mass of 3.5 M$_\odot$ for the secondary and an orbital inclination of 65$^\circ$, the implied primary mass is 0.46 M$_\odot$ - consistent with the system experiencing the CE while the primary was on the RGB.  Models of post-RGB evolution have demonstrated that such systems are capable of ionizing PNe \citep{hall13}, and a handful of other candidate (much shorter orbital period) post-RGB PNe have been discovered \citep{hillwig17}.  If the nebula is the product of a CE on the RGB, then it would be expected to be more massive (as the envelope mass on the RGB is greater than on the AGB).  This is seemingly borne out by estimates of the molecular mass of the nebula performed by \citet{arias01}, who concluded that for a distance of 0.69 kpc the total molecular mass of the nebula was $\sim$0.8 M$_\odot$ which when scaled for to the new distance of 1.45 kpc implies an estimated molecular mass of approximately 3.5 M$_\odot$.  It is important to highlight that this value is extremely dependent on the adopted pre-shock density, with more extreme values resulting in almost an order of magnitude difference in the estimated molecular mass \citep[see section 3.5.2 of][for a full discussion]{arias01} - nonetheless this estimate (derived using an intermediate value of the pre-shock density) serves to show that the PN may, indeed, be the product of a CE while the primary was on the RGB.

In summary, we confirm the central star of NGC~2346 to be one of the longest period post-CE systems known \citep[see e.g.\ ][]{rebassa-mansergas12,sowicka17b,miszalski18}.  Furthermore, the secondary of the system is found to be sub-giant which, in spite of the larger uncertainties involved, must be one of, if not the most, massive post-CE secondaries known \citep{davis10,zorotovic10}.  These properties make the central star of NGC~2346 an important system with which to study the CE and particularly the dependence of the CE efficiency on the parameters of the stellar components, extending the parameter space to longer periods and higher secondary masses \citep[where, intriguingly, an anti-correlation between secondary mass and CE efficiency has been found;][]{davis12}.  Unfortunately, the mass of the primary (another key ingredient in extending the parameter space of these studies) is unknown, however, assuming that the binary orbital plane is at least close to aligned with the waist of the bipolar nebula would imply a low primary mass perhaps consistent with a post-RGB object rather than post-AGB.  Alternatively, NGC~2346 may be the first post-CE system in which the binary plane is not coincident with the nebular waist.

\section*{Acknowledgements}

We thank the anonymous referee for their useful comments. This research was based on observations obtained with the HERMES spectrograph, which is supported by the Fund for Scientific Research of Flanders (FWO), Belgium, the Research Council of KU Leuven, Belgium, the Fonds National de la Recherche Scientifique (FNRS), Belgium, the Royal Observatory of Belgium, the Observatoire de Gen\`eve, Switzerland and the Th\"uringer Landessternwarte Tautenburg, Germany. The Mercator telescope is operated thanks to grant number G.0C31.13 of the FWO under the ``Big Science'' initiative of the Flemish government. H.V.W. acknowledges support from the Research Council of the KU Leuven under grant number C14/17/082. This paper is based on observations made with ESO Telescopes at the La Silla Paranal Observatory under programme IDs 093.D-0038 and 096.D-0080. This research has been supported by the Spanish Ministry of Economy and Competitiveness (MINECO) under the grant AYA2017-83383-P.  The authors wish to thank all observers of the HERMES consortium institutes (KU Leuven, ULB, Royal Observatory, Belgium, and Sternwarte Tautenburg, Germany) who contributed to this monitoring programme.




\bibliographystyle{mnras}
\bibliography{literature} 



\appendix

\section{Radial velocity measurements}
\begin{table}
\caption{Radial velocity measurements of the central star of NGC~2346}
\label{tab:rvs}
\begin{tabular}{lrl}
\hline
Barycentric Julian date & \multicolumn{2}{c}{Barycentric radial velocity}\\
& \multicolumn{2}{c}{(\kms{})}\\
\hline
2457720.75501 & 19.0 & $\pm$1.7 \\
2457720.77642 & 19.0 & $\pm$1.8 \\
2457727.69405 & 39.6 & $\pm$1.9 \\
2457727.71547 & 39.2 & $\pm$1.5 \\
2457736.63952 & 17.2 & $\pm$2.1 \\
2457736.66094 & 17.6 & $\pm$1.9 \\
2457788.60053 & 23.9 & $\pm$1.9 \\
2457788.61847 & 23.5 & $\pm$1.8 \\
2457792.59375 & 42.4 & $\pm$1.8 \\
2457792.61517 & 42.6 & $\pm$1.8 \\
2457801.38023 & 15.2 & $\pm$1.9 \\
2457801.40164 & 14.8 & $\pm$2.2 \\
2457802.36027 & 16.4 & $\pm$2.0 \\
2457802.38168 & 16.2 & $\pm$2.1 \\
2457806.34486 & 32.6 & $\pm$1.8 \\
2457806.36627 & 33.6 & $\pm$1.6 \\
2457807.53482 & 38.8 & $\pm$1.8 \\
2457807.55623 & 39.0 & $\pm$1.8 \\
2457810.39931 & 42.8 & $\pm$1.6 \\
2457810.42072 & 42.8 & $\pm$1.7 \\
2457812.41361 & 39.0 & $\pm$1.7 \\
2457812.43502 & 39.3 & $\pm$1.6 \\
2457813.48179 & 34.1 & $\pm$1.8 \\
2457813.50789 & 33.9 & $\pm$1.9 \\
2457815.40881 & 22.5 & $\pm$1.9 \\
2457815.43022 & 22.8 & $\pm$1.7 \\
2457816.36321 & 17.7 & $\pm$1.8 \\
2457816.38462 & 19.1 & $\pm$1.8 \\
2457857.40471 & 44.0 & $\pm$1.8 \\
2457857.42612 & 44.6 & $\pm$1.9 \\
2458106.75374 & 16.4 & $\pm$2.0 \\
2458227.36508 & 41.7 & $\pm$1.8 \\
2458227.38649 & 41.7 & $\pm$1.7 \\
\hline
\end{tabular}
\end{table}


\bsp	
\label{lastpage}
\end{document}